\documentclass[number,5p]{elsarticle}
\usepackage{textcomp}
\usepackage{graphicx}
\usepackage{amsmath}
\usepackage{amssymb}
\usepackage[usenames]{color}
\biboptions{square,comma,sort&compress}
\usepackage{hyperref}

\journal{Journal of Crystal Growth}

\begin{document}
\begin{frontmatter}

\title{Encapsulated  Single  Crystal   Growth  and  Annealing  of  the
  High-Temperature Superconductor Tl-2201}
\author[fizz]{D.C. Peets\corref{dpeets}\fnref{peets}}
\ead{dpeets@scphys.kyoto-u.ac.jp}
\author[fizz,cifar]{Ruixing Liang}
\author[eos]{Mati Raudsepp}
\author[fizz,cifar]{W.N. Hardy}
\author[fizz,cifar]{D.A. Bonn\fnref{bonn}}
\ead{bonn@physics.ubc.ca}
\address[fizz]{University of British Columbia, Department of Physics
  \& Astronomy, 6224 Agricultural Road, Vancouver, BC, Canada V6T 1Z1}
\address[eos]{University of British Columbia, Department of Earth 
  \& Ocean Sciences, 6339 Stores Road, Vancouver, BC, Canada V6T 1Z4}
\address[cifar]{Canadian Institute for Advanced Research, Canada}
\cortext[dpeets]{Corresponding author}
\fntext[peets]{Current address: Department of Physics, Graduate School
  of   Science,  Kyoto   University,  Kyoto   606-8502,   Japan,  Tel:
  +81-75-753-3744, FAX: +81-75-753-3783}
\fntext[bonn]{Requests for samples should be directed to D.A.~Bonn.}

\begin{abstract}

Highly-perfect platelet single crystals of Tl$_2$Ba$_2$CuO$_{6+\delta}$ 
(Tl-2201) were grown by a self-flux technique.  A novel encapsulation 
scheme allowed the precursors to react prior to the sealing required to 
contain volatile thallium oxides, and permitted the removal of melt at 
the conclusion of growth, reproducibly producing high yields of clean 
crystals.  The crystals were annealed under well-controlled oxygen 
partial pressures, then characterised.  They have sharp superconducting 
transitions, narrow X-ray rocking curves and a low 4\% substitution of 
thallium by copper, all evidence of their high perfection and 
homogeneity.  The crystals are orthorhombic at most dopings, and a 
previously unreported commensurate superlattice distortion is observed.

\end{abstract}

\end{frontmatter}

\section{Introduction}\label{sec:intro}

In condensed matter  physics, a great deal of  effort is being applied
to  the  problem  of  correlated electron  systems,  and  particularly
high-temperature  superconductivity (HTSC)  in the  cuprates.  Because
these superconductors are highly anisotropic and because line nodes in
the  superconducting  energy  gap  make  them  highly  susceptible  to
impurities, accurate measurements  of their electronic properties rely
crucially on  the availability  of highly perfect,  high-purity single
crystals.

In  the  cuprates,  hole  doping  accesses a  progression  of  unusual
electronic  phases; however, at  high doping  levels (above  where the
critical temperature  $T_c$ peaks) there  is evidence for a  return to
the relative normalcy of  Fermi liquid theory, including a resistivity
approaching $T^2$ \cite{Mackenzie1996}  and a Fermi surface resembling
that      predicted      by      band      structure      calculations
\cite{Hussey2003,Plate2005,Peets2007}.     Despite   the   tantalising
prospect of a  phase that can be readily  understood, this `overdoped'
regime where $T_c$ falls back  to zero has received significantly less
attention  than  lower doping  ranges,  due  largely  to a  dearth  of
suitable samples.

\sloppy
One      particularly     promising     overdoped      material     is
Tl$_2$Ba$_2$CuO$_{6+\delta}$  (Tl-2201) \cite{Sheng1988}, which  has a
particularly flat CuO$_2$ plane, a relatively simple crystal structure
containing no  (CuO$_2$)$_n$ multilayers or CuO chain  layers, and can
be overdoped  to $T_c=0$.  Besides  being suitable for  bulk transport
measurements,  Tl-2201  has  a  non-polar cleavage  plane  within  its
Tl$_2$O$_2$   double  layer   that   ensures  that   surface-sensitive
single-particle  spectroscopies provide information  characteristic of
the bulk.  Indeed,  Tl-2201 recently became the first  HTSC cuprate on
which bulk and surface  measurements agreed quantitatively on the same
physical  property  ---  the   Fermi  surfaces  measured  via  angular
magnetoresistance    oscillations    (AMRO)   \cite{Hussey2003}    and
angle-resolved  photoemission  spectroscopy \cite{Plate2005,Peets2007}
(ARPES).  The excellent agreement  indicated that Tl-2201 may be ideal
for finally  joining the modern single-particle  spectroscopies with a
host of well-established bulk probes.

\fussy
Crystals  of   Tl-2201  have   been  grown  previously   by  self-flux
\cite{Liu1992,Kolesnikov1995,Hasegawa2001}      and      KCl      flux
\cite{Manako1992} techniques.   The chief complication  in this system
is the  formation of monovalent  Tl$_2$O, which has a  vapour pressure
around 0.2--0.3~atm  at the growth  temperature \cite{Siegal1997}.  To
avoid  loss  of  Tl$_2$O  the  system must  be  sealed;  however,  the
conversion of Tl$_2$O$_3$ to Tl$_2$O also produces oxygen, which leads
to high system pressure if not exhausted.  While most groups attempted
to  limit   thallium  loss   by  impeding  diffusion,   only  Hasegawa
\cite{Hasegawa2001}  reported  an  encapsulation scheme  to  {\slshape
  prevent}  loss of thallium.   {\sl In  situ} separation  of crystals
from  high-temperature  melt  has  not been  reported;  crystals  were
typically separated mechanically from cooled, solidified flux, leading
to low crystal yields.  In the case of KCl flux growth, the separation
is  relatively straightforward,  but the  crystals produced  are quite
small and there is a risk of potassium contamination.

Tl-2201's hole doping (and thus  its $T_c$) is controlled via variable
occupancy of an oxygen interstitial between the Tl--O layers.  Setting
the oxygen  content to desired  levels requires post-annealing  of the
crystals at controlled temperatures and oxygen partial pressures.  The
vapour pressure of Tl  at elevated temperatures makes proper annealing
procedures essential to obtain  high-quality crystals, because loss of
Tl   means    degradation   and   decomposition    of   the   crystals
\cite{Vyaselev1992}.  Here we report novel processes we have developed
for  self-flux  growth and  post-growth  annealing  of Tl-2201  single
crystals.


\section{Single Crystal Growth and Annealing}\label{sec:growth}

Single  crystals of  Tl-2201 were  grown  by a  copper-rich self  flux
method  in encapsulated  crucibles.   Precursor powders  of CuO  (Alfa
Aesar,  99.995\%), BaO$_2$  (Aldrich, 95\%,  major  impurity BaCO$_3$,
heavy  metals 0.01\%,  50~ppm Fe)  and Tl$_2$O$_3$  (Aldrich, 99.99\%)
were   intimately  mixed   with  the   cation  ratio   Tl~:~Ba~:~Cu  =
2.2~:~2~:~1.8,  then packed  into  an alumina  crucible containing  an
upright  gold  sieve.   Barium  carbonate  was not  used  because  its
decomposition  to   BaO  is  very  slow  at   this  system's  liquidus
temperature.

\begin{figure}[htbp]
\begin{center}
\includegraphics[width=0.5\columnwidth]{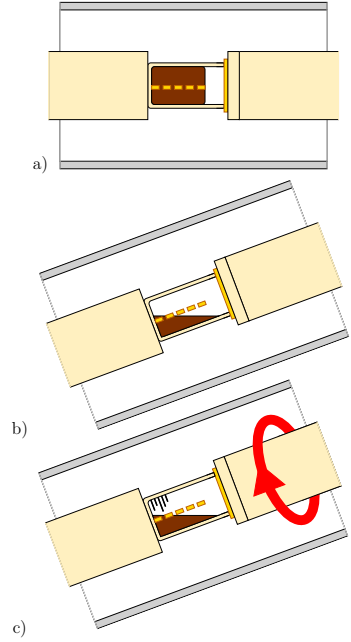}
\caption{\label{fig:decant}(Colour online)  Schematic depiction of the
  {\sl  in situ} flux  separation scheme.   {\bfseries a)}  The filled
  crucible and  gold lid  are inserted between  two alumina  rams; the
  crucible contains an upright gold sieve.  {\bfseries b)} The furnace
  is  inclined, the  charge melted,  and  crystals are  grown on  slow
  cooling.   {\bfseries  c)}  The  rams are  rotated,  separating  the
  crystals from excess flux with the aid of the gold sieve.}
\end{center}
\end{figure}

The growth technique is depicted in Fig.~\ref{fig:decant}.  A gold lid
was  affixed to  the crucible,  then the  crucible and  lid  were held
between two  alumina rams  in a tube  furnace inclined  20$^\circ$ and
supporting a transverse temperature gradient.  Sufficient pressure was
applied to the rams to hold the crucible in place, but not to seal it,
while the crucible was heated  and the precursors reacted.  This acted
as a barrier  to the loss of thallium by  diffusion while allowing the
release of any  overpressure due to evolved oxygen.   After soaking at
935$^\circ$C (above the liquidus) for  two hours to allow evolving gas
to  exhaust, the pressure  applied to  the gold  lid was  increased to
several atmospheres, sealing the crucible.  Evolved vapours were swept
from  the furnace  by flowing  oxygen,  then bubbled  through acid  to
remove any thallium.

The melt's temperature was reduced  at a rate of -0.3$^\circ$C/h along
the  liquidus curve  to 890$^\circ$C,  just above  the Tl-2201  -- CuO
eutectic,  which  was  found  to  be slightly  lower  than  previously
reported  \cite{Jorda1993}.  At  this temperature,  the  crystals were
separated from the remaining melt by rotating the rams and crucible by
180$^\circ$; the  gold sieve ensured that the  crystals remained clear
of the  melt.  The  temperature was reduced  to 825$^\circ$C for  a 48
hour anneal to make the  distribution of substituted copper atoms more
homogeneous, followed by free cooling to room temperature.

\sloppy
The charge's mass was measured  before and after growth; the mass lost
was typically less than that  expected assuming the loss of one oxygen
atom from BaO$_2$ and  two from Tl$_2$O$_3$.  Since BaO$_2$ decomposes
around 800$^\circ$C  and Tl$_2$O$_3$ at higher  temperatures, an upper
limit of  5\% may be obtained for  the thallium lost if  it is assumed
that all  thallium remains  trivalent and all  mass losses  beyond the
decomposition of  BaO$_2$ are attributable to loss  of thallium oxide.
Hasegawa {\it et  al.}  \cite{Hasegawa2001} pre-reacted the precursors
below the  eutectic temperature  prior to encapsulating  the crucible,
but some crucibles still  ruptured from the pressure, suggesting gases
are  evolved  when  the  precursors  melt.   Here,  the  crucible  was
encapsulated  after  the  melting  of  the  precursors.   This  avoids
eruption  of  the  crucible  contents  and allows  for  more  reliable
sealing, but at the cost of a few percent of the thallium.

\fussy
Irregularly   shaped  black,   platelet   single  crystals   typically
$1\times1\times0.01$~mm$^2$  in size with  mirror surfaces  were grown
reproducibly  by this  method.  After  mechanical separation  from the
crucible, they were annealed under controlled oxygen partial pressures
at temperatures between 290$^\circ$C and 500$^\circ$C, producing sharp
superconducting transitions  at temperatures  ranging from 5  to 85~K.
No  attempts  to  reach  optimal  doping (corresponding  to  at  least
$T_c=93$~K   \cite{Wagner1997})   were    made,   to   avoid   surface
decomposition, but  some batches exhibited as-grown $T_c$s  as high as
90~K.  The  $T_c$ produced by  the oxygen partial pressure  and anneal
temperature   resembled   those   reported  previously   on   ceramics
\cite{Opagiste1993a}.

\begin{figure}[htbp]
\begin{center}
\raisebox{4pt}{a)}
\includegraphics[width=0.75\columnwidth]{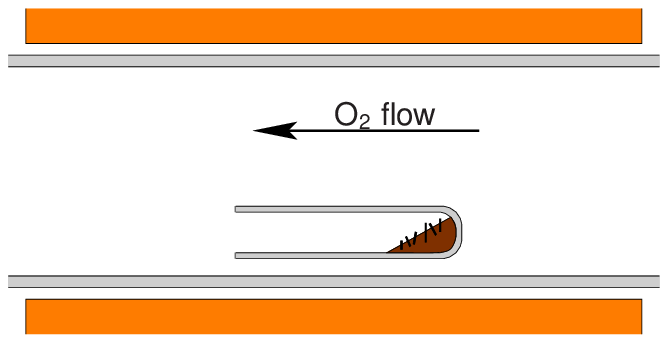}\\
\vspace*{12pt}
\raisebox{4pt}{b)}
\includegraphics[width=0.75\columnwidth]{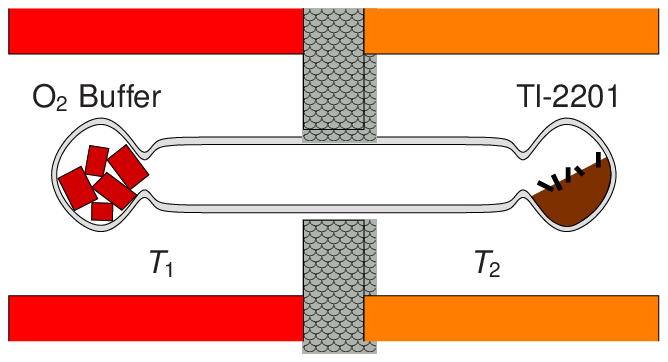}
\caption{\label{fig:anneal}(Colour  online) Annealing schemes  used to
  obtain a)  strongly overdoped and b)  moderately overdoped crystals.
  In a), the  crystals are placed on Tl-2201  powder in flowing oxygen
  or an appropriate  mixed gas; in b), the Tl-2201  powder serves as a
  thallium buffer,  an oxygen partial  pressure buffer is  situated at
  the other end  of the quartz ampoule and  the two ends' temperatures
  may be controlled separately for independent control over the oxygen
  partial pressure and the anneal temperature.}
\end{center}
\end{figure}

Two oxygen-annealing schemes are depicted in Fig.~\ref{fig:anneal}. To
access   the  strongly  overdoped   regime  (high   oxygen  contents),
corresponding to $T_c$s below  $\sim45$~K, oxygen partial pressures of
$10^{-4}$ to  1~atm were  applied via flowing  mixtures of  oxygen and
nitrogen gases.   Within this oxygen partial pressure  range, the loss
of thallium from the crystals remained undetectable even after anneals
of several weeks.

To  access the  lower  doping regime,  corresponding  to $T_c$s  above
$\sim45$~K,   annealing  at  oxygen   partial  pressures   lower  than
$10^{-4}$~atm is  required \cite{Opagiste1993a}.  At  these low oxygen
pressures, the loss of thallium from the crystals becomes an issue and
the  annealing must  be carried  out in  an enclosed  environment.  As
shown  in  Fig.~\ref{fig:anneal}(b),   Tl-2201  crystals  embedded  in
Tl-2201 powder were  placed in one end of a  sealed quartz capsule and
an oxygen  buffer consisting of  a mixture of  CuO and Cu$_2$O  at the
other end.   The two ends  were thermally isolated by  fibrous ceramic
insulation so  that the  temperatures of the  crystals and  the buffer
could  be  controlled  independently.   The  Tl-2201  powder  releases
sufficient Tl$_2$O to produce the required equilibrium Tl$_2$O partial
pressure in the  capsule, essentially eliminating the loss  of Tl from
the  crystals.  The  oxygen  partial pressure  is  controlled via  the
temperature of  the buffer,  through the chemical  equilibrium between
CuO and Cu$_2$O.


At  the conclusion  of  each  anneal, the  crystals  were quenched  by
plunging  the quartz  tube or  capsule  into an  icewater bath.   This
preserves  the  equilibrium   oxygen  content  established  under  the
annealing conditions and ensures a homogeneous dopant distribution.

\section{Characterisation}

\sloppy
X-ray rocking curves  of several Tl-2201 crystals were  collected on a
Philips X'Pert Pro single crystal X-ray dif\-fractometer, using a copper
{\slshape  K}$\alpha_1$  vertical line  source  excited  by 45~kV  and
40~mA.  Diffracted  X-rays were detected using a  serial detector with
no scanning slit.  Each  $\sim1$~mm$^2$ crystal was fully illuminated.
Rocking  curve  widths  (FWHM)  were $0.025^\circ  \sim  0.050^\circ$,
indicating a high degree of crystalline perfection --- comparable to
good  YBa$_2$Cu$_3$O$_{6+\delta}$  grown  in  zirconia  crucibles.   A
typical      (0~0~10)     rocking      curve      is     shown      in
Fig.~\ref{fig:spots}(a).

\fussy
Electron-probe   micro-analyses  (EPMA)   of  several   crystals  were
performed   using  a  fully-automated   CAMECA  SX-50   instrument  in
wavelength-dispersion  mode.  EPMA was  performed using  an excitation
voltage of 15~kV,  a beam current of 20~nA,  peak and background count
times of 80  and 40~s respectively, and a  spot diameter of 10~$\mu$m;
initial data  reduction employed the ``PAP''  $\varphi(\rho Z)$ method
\cite{PAP}.  For the cations  analyzed, the following standards, X-ray
lines and monochromator crystals  were used: elemental Tl, Tl{\slshape
  M}$\alpha$,  PET; YBa$_2$Cu$_3$O$_{6.920}$,  Ba{\slshape L}$\alpha$,
PET; and YBa$_2$Cu$_3$O$_{6.920}$, Cu{\slshape K}$\alpha$, LIF.  Tight
Pulse  Height Analysis  (PHA) control  was  used to  eliminate to  the
degree possible any interference from higher-order lines.

Crystals were epoxied  to the face of an acrylic  disc; care was taken
to ensure the  surface remained parallel to the  disc, and 10--15 flat
sites  were  studied  per  crystal,  then  averaged.   No  significant
variations were seen among sites on the same crystal or among crystals
from  similar growth  conditions, but  variations were  observed among
crystals  grown using  different initial  cation ratios.   Because the
barium  results   were  less  consistent,  possibly   due  to  surface
topography, the cation composition was normalised to (Tl~+~Cu)~=~3 ---
in Tl-2201, Cu  is known to substitute for Tl  and no cation vacancies
exist.     The    cation   composition    was    determined   to    be
Tl$_{1.920(2)}$Ba$_{1.96(2)}$Cu$_{1.080(2)}$O$_{6+\delta}$   (2$\sigma$
uncertainties),  indicating  that 4\%  of  Tl  is  substituted by  Cu,
slightly       lower      than       other       reported      results
\cite{Shimakawa1990,Liu1992,Kolesnikov1992},    including    the   5\%
substitution   in  the   author's  earlier   generation   of  crystals
\cite{Peets2007}.  Better  encapsulation, and  hence lower loss  of Tl
during  the growth,  may have  contributed to  the lower  level  of Cu
substitution for Tl.  A separate check for aluminum contamination from
the crucible excluded  it at the 50~ppm level; it  was not possible to
check for gold  contamination, but this is expected  to be minimal due
to  the absence  of sites  like YBCO's  CuO chains  that are  known to
accommodate Au$^{1+}$.

\begin{figure}[htbp]
\begin{center}
\includegraphics[width=\columnwidth]{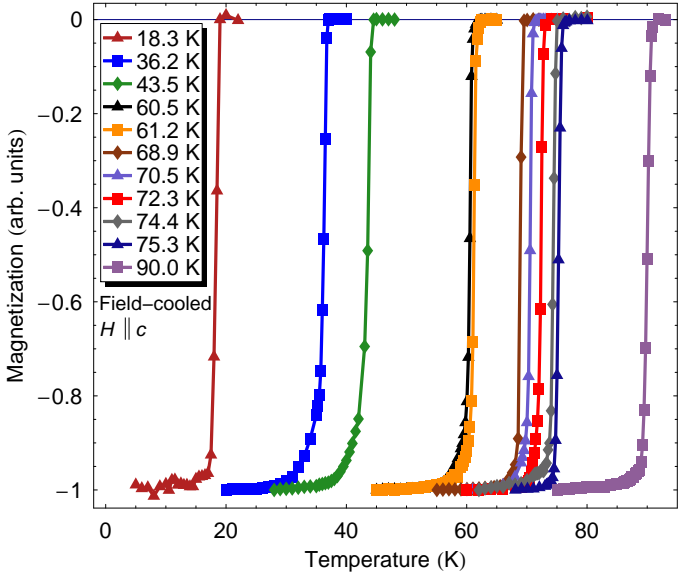}
\caption{\label{fig:squid}(Colour  online)  Field-cooled magnetisation
  curves of several Tl-2201 crystals.  The $T_c = 90$~K crystal was as
  grown; the  others were annealed.  Transition widths  (10\% -- 90\%)
  are typically $0.75\sim1.5$~K ---  the crystals may be homogeneously
  doped  over a  wide doping  range.   Data were  collected in  fields
  $H\parallel c$ of $1\sim2$~Oe.}
\end{center}
\end{figure}

The crystals'  transition temperatures  $T_c$ and widths  $\Delta T_c$
were  characterised using  a Quantum  Design MPMS  SQUID magnetometer;
fields $H\parallel  c$ of only  $1\sim2$~Oe were employed  to minimise
broadening.   Normalised field-cooled  SQUID magnetisation  curves for
Tl-2201   crystals   at   a   variety   of  dopings   are   shown   in
Fig.~\ref{fig:squid}.  Transition widths  of $1\sim2$~K, indicative of
homogeneous doping,  can be obtained throughout the  doping range from
18~K  to 75~K.   These transitions  are narrower  than  those reported
previously  on  Tl-2201,  and  compare  well with  crystals  of  other
cuprates.  The  Meissner fraction (the  fraction of the  magnetic flux
excluded when cooling through $T_c$) was typically $40\sim60$\%, which
compares well  with field-cooled  results on high-quality  crystals of
other cuprates.

Table \ref{tab:new}  compares the  results of previous  single crystal
studies   with  the  abovementioned   techniques  used   for  thallium
containment  and  annealing,  selected transition  temperature  widths
$\Delta T_c$,  and several crystal structure  results discussed below.
The transition widths shown  in Fig.~\ref{fig:squid} compare well with
previous results,  indicating the high quality and  homogeneity of the
crystals.

\begin{table*}[htbp]
\caption{\label{tab:new}Comparison to literature:  the preparation and properties of the crystals reported in this work are compared with selections from previous reports on single-crystalline Tl-2201.  Included are containment method and annealing performed; $T_c$ with its width $\Delta T_c$ reported as an uncertainty in the last digit; cation substitution level $z$ and oxygen content $6+\delta$ in (Tl$_{1-z}$Cu$_z$)$_2$Ba$_2$CuO$_{6+\delta}$ by EPMA and X-ray diffraction; and space group.  A dash indicates properties that were not reported; crystals on which X-ray refinement was not performed are denoted ``tetragonal'' and ``orthorhombic'' in place of a space group.  The crystal reported in Ref.\ \cite{Kolesnikov1992} was grown in the presence of calcium.}
\begin{center}
\begin{tabular}{|l|c|c|c|c|c|c|c|} \hline
Source & Sealing & Annealing & $T_c$ ($\Delta T_c$) & $z$ (EPMA) & $z$ (X-ray) & $6+\delta$ (X-ray) & Symmetry\\ \hline\hline
This work & Au lid, & Cation, & 90(1) K & --- & --- & --- & Tetragonal \\ \cline{4-8}
& variable & $P_{\text{O}_2}/P_{\text{Tl}_2\text{O}}$ & 75(1) K & 0.040 & 0.043 & 6.29(18) & Fmmm \\ \cline{4-8}
& pressure & & 55(2) K & 0.040 & --- & --- & Orthorhombic \\ \cline{4-8}
& & & 9(5) K & --- & --- & --- & Orthorhombic \\ \hline
Ref.\ \cite{Torardi1988} & Au tube & --- & 90(7) K & --- & --- & --- & I4/mmm \\ \hline
Ref.\ \cite{Liu1992} & Au lid & $P_{\text{O}_2}$, & 12.4(8) K & 0.075 & 0.068 & 6.0 & I4/mmm \\
& & see \cite{Mackenzie1996} & & & & & \\ \hline
Ref.\ \cite{Kolesnikov1992} & Alunde anvil & --- & 110(15) K & 0.05 & 0.073 & 6.0 & I4/mmm \\ \hline
Ref.\ \cite{Kolesnikov1995} & Multilayered & --- & 30(---) K & --- & 0.050 & 6.18(4) & I4/mmm \\
& crucible & & & & & & \\ \hline
Ref.\ \cite{Hasegawa2001} & Au capsule, & $P_{\text{O}_2}$ & 25(3) K & Vacancy: & --- & --- & I4/mmm \\
& Al$_2$O$_3$ bomb & {\it in situ} & & 0.05 & & & \\ \hline
Ref.\ \cite{Peets2007} & Weighted & $P_{\text{O}_2}$, no & 24(4) K & 0.055 & --- & --- & Orthorhombic \\ \cline{4-8}
& Au lid & details & 67.7(7) K & 0.055 & --- & --- & Tetragonal \\ \hline
\end{tabular}
\end{center}
\end{table*}

\section{Crystal Structure}\label{sec:struct}

X-ray and neutron powder  diffraction studies on Tl-2201 have revealed
the  existence   of  two  distinguishable  phases   of  Tl-2201,  with
orthorhombic             and            tetragonal            symmetry
\cite{Shimakawa1989,Shimakawa1993}.     Both   the   degree    of   Cu
substitution  on the  Tl site  and the  amount of  interstitial oxygen
between     the     Tl--O     layers     control     the     structure
\cite{Torardi1988,Liu1992,Kolesnikov1992,Strom1994,Aranda1995,Wagner1997}.
Oxygen  vacancies in  the Tl--O  layers  have also  been observed  for
samples  with low  oxygen contents  (near optimal  doping).   The most
thorough  structure   investigation  was   by  Wagner  {\it   et  al.}
\cite{Wagner1997},  which  elucidated  the  interplay  between  oxygen
interstitials,  oxygen  vacancies,  cation  substitution,  $T_c$,  and
orthorhombicity.  Structure  refinement by single  crystal diffraction
should  provide more  structural details,  but so  far  single crystal
X-ray diffraction \cite{Liu1992,Kolesnikov1992}  has only been carried
out on tetragonal crystals;  the high concentration of Cu substituents
and low  occupancy of interstitials may make  some structural features
difficult to  observe.  Measurements of the lattice  parameters on the
author's  earlier crystals  showed evidence  of  orthorhombic symmetry
\cite{Peets2007},  not  previously reported  in  crystals.  Since  the
orthorhombicity was  enhanced at high  oxygen contents, the  choice of
near-optimally doped crystals may be responsible for previous studies'
non-observation of orthorhombicity.  However, earlier work has equated
orthorhombicity   with  stoichiometric,   non-superconducting  Tl-2201
\cite{Hewat1988,Shimakawa1993}, suggestive of  a link between symmetry
and superconductivity,  and orthorhombicity has only  been reported in
one  crystal.   Here  we  provide  confirmation  that  superconducting
crystals may be orthorhombic as well as tetragonal.


Lattice parameters were measured on several dopings of Tl-2201 crystal
at -100$^\circ$C on a Bruker  X8 Apex diffractometer, using a CCD area
detector  and  a molybdenum  X-ray  source.   A  full X-ray  structure
refinement  was performed  on a  crystal with  $T_c=75$~K,  using 3319
reflections  (312 unique)  in  the $k$-space  region $-7\leq  h\leq7$,
$-7\leq  k\leq7$, $-31\leq  l\leq32$, by  a  full-matrix least-squares
minimization  of  $F^2$.   Structure  refinement was  performed  using
SHELXL-97  \cite{ShelX},  in the  orthorhombic  space  group Fmmm  for
generality;  this unit  cell  is rotated  45$^\circ$  relative to  the
tetragonal  cell and  contains four  formula units.   All  atoms were
constrained to their symmetric positions (relaxing this constraint led
to only minor  improvements), no sites were split,  and occupancies of
all sites except  O(4) were fixed at full  occupancy; atoms other than
O(4)  were allowed anisotropic  thermal parameters.   Besides refining
for occupancy of the O(4)  site, a refinement was performed for copper
substitution  on the thallium  site.  Note  that since  all previously
published data  were collected at room  temperature, thermal expansion
must  be taken  into account  when comparing  these data  to published
values.

The    crystals'     lattice    parameters    are     summarised    in
Table~\ref{tab:lattice};  orthorhombicity is  characterised  using the
orthorhombic strain $\eta  = 2(b-a)/(b+a)$ as a figure  of merit.  The
$T_c=9$~K crystal was extracted from an early batch that leaked during
growth, likely leading to higher cation substitution.  Aside from this
crystal,  more  heavily  overdoped  samples (lower  $T_c$s)  are  more
orthorhombic.     This   is    consistent   with    Wagner's   results
\cite{Wagner1997},   and   constitutes   important   confirmation   of
orthorhombicity in superconducting  single crystalline Tl-2201 --- the
symmetry does not determine whether the material superconducts.  

\begin{table}[htbp]
\begin{center}
\caption{\label{tab:lattice}Lattice parameters and orthorhombic strain
  $\eta$ for  four dopings of Tl-2201 as  determined by single-crystal
  X-ray  diffraction;  uncertainties  are  $1\sigma$.   The  $T_c=9$~K
  crystal is thought to have higher cation substitution.}
\begin{tabular}{|r|r@{.}l|r@{.}l|r@{.}l|r@{.}l|}\hline
$T_c$ & \multicolumn{2}{c|}{$a$ (\AA)} & \multicolumn{2}{c|}{$b$ (\AA)}
& \multicolumn{2}{c|}{$c$ (\AA)} & \multicolumn{2}{c|}{$\eta$ 
(\textperthousand)}\\ \hline \hline
90~K & 5&4603(14) & 5&4602(13) & 23&1901(59) & 0&0(4)\\ \hline
75~K & 5&4477(9) & 5&4484(9) & 23&1711(35) & 0&13(23)\\ \hline
55~K & 5&4424(12) & 5&4550(12) & 23&1428(51) & 2&31(31)\\ \hline
9~K & 5&4067(15) & 5&4109(14) & 22&9219(54) & 0&77(35)\\ \hline
\end{tabular}
\end{center}
\end{table}

\begin{figure}[htbp]
\begin{center}
\includegraphics[width=0.84\columnwidth]{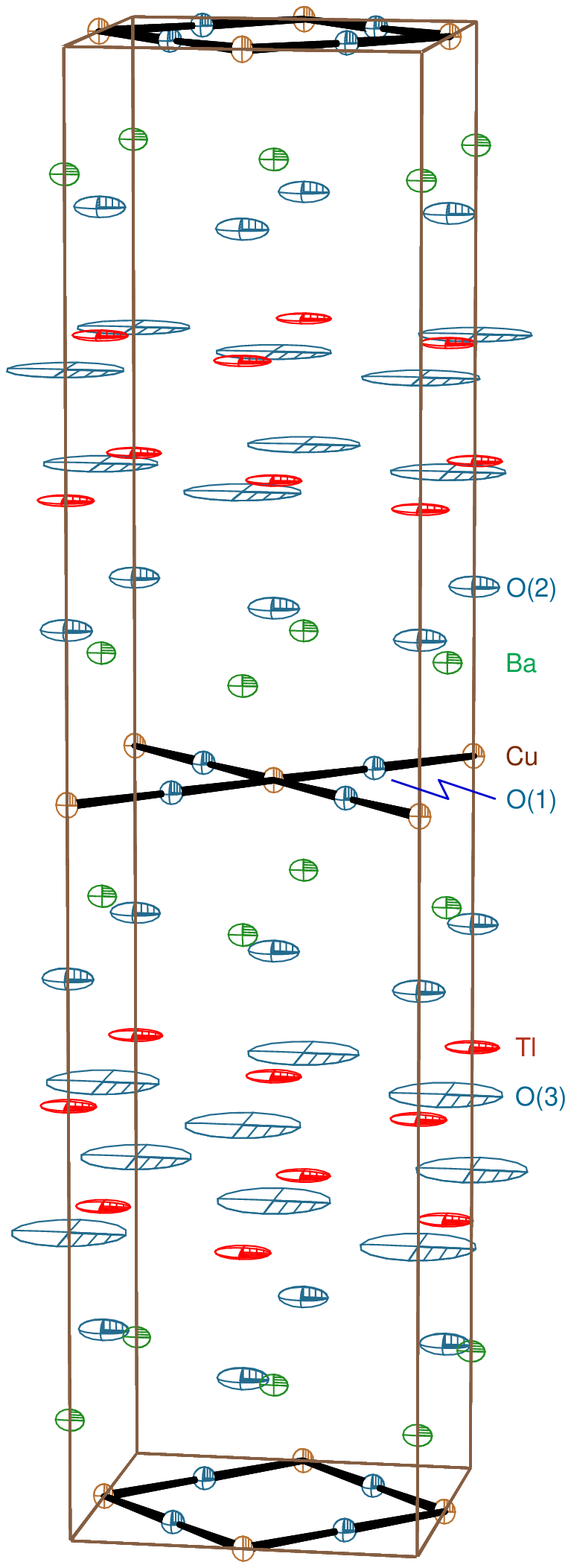}
\caption{\label{fig:ellipsoid}(Colour online)  Refined orthorhombic
  crystal structure  with 98\% probability ellipsoids  for each atom's
  nuclear position, $T_c=75$~K Tl-2201 crystal, indentifying each site
  and clearly  suggesting that  the structure of  the Tl--O  and Ba--O
  layers  is   not  fully  captured  by  the   refinement.   The  O(4)
  interstitial  has been  excluded for  clarity.}
\end{center}
\end{figure}

The refined atomic positions  for the $T_c=75$~K crystal are presented
in Table~\ref{tab:XRD_xyz},  the full thermal  displacement parameters
$U_{ij}$  for these atoms  are reproduced  in Table~\ref{tab:XRD_Uij},
and  the   resulting  orthorhombic  crystal  structure   is  shown  in
Fig.~\ref{fig:ellipsoid}.  Because no sites were split, there were few
refinable  parameters.   The substitution  of  copper  atoms onto  the
thallium site was refined with the site constrained to full occupancy,
yielding  the substitution level  Tl$_{1.914(14)}$Cu$_{0.086(14)}$, in
excellent agreement with the EPMA result.  The O(4) occupancy was also
refined, but the low sensitivity  of this technique to light atoms and
the  low occupancy  of the  site at  this doping  led to  an occupancy
consistent with  zero to within  2$\sigma$ --- O$_{0.29(18)}$  --- and
the  site  departed  from   its  previously  reported  position.   The
refinement's $R_1$ factor  was 2.35\%; $wR_2$ was 5.68\%  on all data.
The  cation substitution and  O(4) occupancy  results, along  with the
symmetry, are compared against  previous reports on single crystals in
Table \ref{tab:new}.

\begin{table}[htbp]
\begin{center}
\caption{\label{tab:XRD_xyz}Refined  atomic  positions and  equivalent
  isotropic  thermal displacement parameter  $U_{eq}$, defined  as one
  third of  the trace of  the orthogonalised $U_{ij}$ tensor,  for the
  $T_c=75$~K  crystal.  See  Table~\ref{tab:XRD_Uij} for  the $U_{ij}$
  parameters and Fig.~\ref{fig:ellipsoid} for the crystal structure.}
\vspace*{12pt}
\begin{tabular}{|l|r@{.}l|r@{.}l|r@{.}l|r@{.}l|}\hline
Atom & \multicolumn{2}{c|}{$x/a$} & \multicolumn{2}{c|}{$y/b$} &
\multicolumn{2}{c|}{$z/c$} & \multicolumn{2}{c|}{$U_{eq}$}\\ 
\hline\hline
Tl & 0&0000 & 0&0000 & 0&29732(2) & 0&0108(2)\\ \hline 
Ba & 0&0000 & 0&0000 & 0&08306(3) & 0&0028(3)\\ \hline 
Cu & 0&0000 & 0&0000 & 0&5000 & 0&0023(4)\\ \hline
O(1) & 0&2500 & 0&2500 & 0&0000 & 0&0043(13)\\ \hline
O(2) & 0&0000 & 0&0000 & 0&3833(4) & 0&0087(16)\\ \hline
O(3) & 0&0000 & 0&0000 & 0&2112(5) & 0&055(6)\\ \hline
O(4) & 0&2500 & 0&2500 & 0&283(14) & 0&14(15)\\ \hline
\end{tabular}
\end{center}
\end{table}

\begin{table}[htbp]
\begin{center}
\caption{\label{tab:XRD_Uij}Anisotropic  displacement  parameters  for
  those atomic positions in Table~\ref{tab:XRD_xyz} which were refined
  assuming  an anisotropic electron  density.  $U_{23}$  and $U_{13}$,
  zero by symmetry, are excluded  from this table.  Atoms in the Tl--O
  and Ba--O layers are particularly anisotropic, suggesting this to be
  the location of the superlattice modulation.}
\vspace*{12pt}
\begin{tabular}{|l|r@{.}l|r@{.}l|r@{.}l|c|}\hline
Atom & \multicolumn{2}{c|}{$U_{11}$} & \multicolumn{2}{c|}{$U_{22}$} &
\multicolumn{2}{c|}{$U_{33}$} & $U_{12}$\\ \hline\hline
Tl & 0&0130(3) & 0&0190(4) & 0&0006(3) & 0\\ \hline
Ba & 0&0000(4) & 0&0052(4) & 0&0031(4) & 0\\ \hline
Cu & 0&0000(8) & 0&0030(8) & 0&0040(9) & 0\\ \hline
O(1) & 0&000(3) & 0&006(3) & 0&007(3) & 0.004(2)\\ \hline
O(2) & 0&006(4) & 0&017(4) & 0&003(3) & 0\\ \hline
O(3) & 0&086(15) & 0&080(14) & 0&000(4) & 0\\ \hline
\end{tabular}
\end{center}
\end{table}

\begin{figure}[htbp]
\begin{center}
\includegraphics[width=\columnwidth]{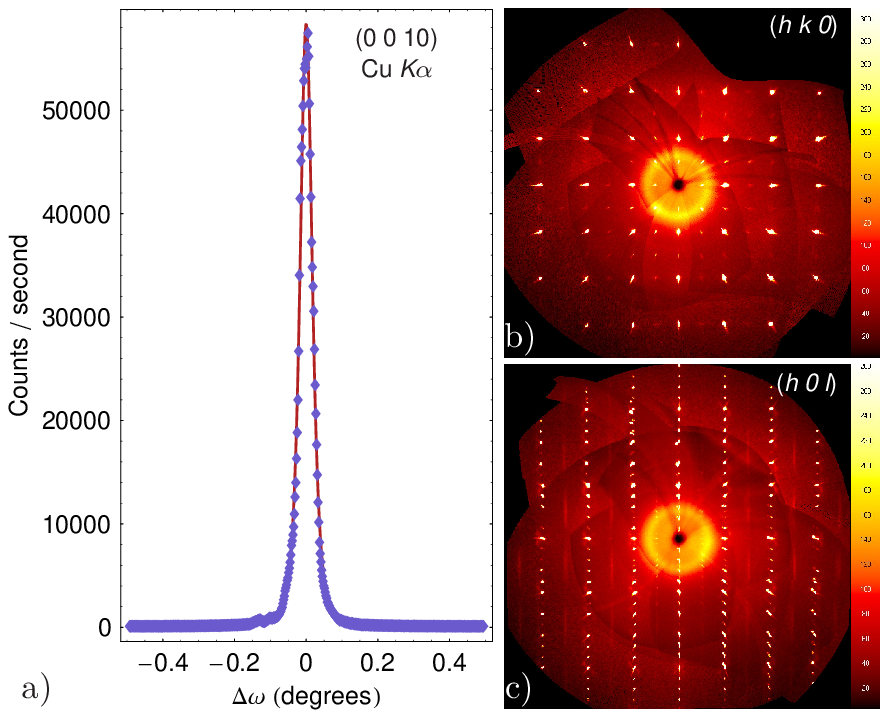}
\caption{\label{fig:spots}(Colour online)  a) (0  0 10)  X-ray
  rocking curve,  $T_c=75$~K crystal.  The full  width at half-maximum
  is  0.040$^\circ$.   b) and  c)  Simulated  X-ray precession  photos
  constructed from the single  crystal diffraction data, for ($h~k~0$)
  and ($h~0~l$)  planes respectively, showing weak  rods of scattering
  corresponding to a commensurate superlattice modulation that doubles
  the unit  cell along $a$ and  $b$ but is  largely uncorrelated along
  $c$.}
\end{center}
\end{figure}

It     is     evident     from     the    crystal     structure     in
Fig.~\ref{fig:ellipsoid} and the  atomic displacement parameters in
Table~\ref{tab:XRD_Uij} that the refinement fails to fully capture the
structure in  the Ba--O and Tl--O  layers.  This can be  dealt with by
splitting         the         Tl         and        O(3)         sites
\cite{Hewat1988,Shimakawa1990,Kolesnikov1992,Wagner1997},         which
provides limited information about  which directions each site departs
from  its  symmetric  position,  while some  studies  have  identified
satellite  peaks  corresponding   to  an  incommensurate  superlattice
modulation            in           the            TlO           layers
\cite{Beyers1988,Shimakawa1993,Aranda1995}.     Simulated   precession
photographs were generated from  the diffraction data, to help clarify
the  crystal  structure.  The  ($h~k~0$)  plane  for the  $T_c$~=~75~K
crystal  is shown  in  Fig.~\ref{fig:spots}(b).  Weak  diffraction
spots  correspond  to  a  commensurate  superlattice  modulation  that
doubles  the in-plane  lattice constants.   As  can be  seen from  the
($h~0~l$)  plane in  Fig.~\ref{fig:spots}(c) (the  ($0~k~l$) plane
appears  similar  and is  not  reproduced  here),  these more  closely
resemble rods  than points, although  they do exhibit  some point-like
character --- the superlattice modulation is not well correlated along
the $c$-axis.

It  is  clear  from  the  thermal  displacement  parameters  that  the
superlattice modulation involves both the  BaO and TlO layers, and the
diffraction patterns indicate that the modulation is commensurate with
the  lattice,  but  the  exact   nature  of  the  modulation  was  not
determined.  That it  is commensurate with the lattice  is a departure
from  previously reported superlattice  modulations in  this material.
The doubling  of the unit cell along  $a$ and $b$ may  be important to
consider in the interpretation of single-particle spectroscopic data.

\section{Conclusion}

\sloppy
In summary,  single crystals of  Tl-2201 were reproduc\-i\-bly grown  by a
copper-rich self-flux  method, employing a  novel encapsulation scheme
that permits the escape of evolved gases as the precursors react, then
seals the crucible for a fully encapsulated growth.  The crystals were
separated from  the molten flux  by rotating the crucible  and letting
the  flux flow through  a gold  sieve.  This  resulted in  much higher
yields  of clean  crystals compared  to the  mechanical  separation of
crystals from solidified flux reported in previous work.  The crystals
were annealed under well-controlled oxygen partial pressures, allowing
the preparation of crystals throughout a wide range of hole doping and
$T_c$.  In particular, crystals with low oxygen contents (high $T_c$s)
were  prepared by  annealing in  an encapsulated  environment  using a
CuO/Cu$_2$O mixture as an oxygen  buffer.  This avoided the loss of Tl
and surface degradation reported in earlier work.

X-ray rocking curves  and magnetisation measurements demonstrated that
the crystals had a very high degree of crystalline perfection and that
the  dopant distribution was  homogeneous.  The  compositions obtained
from   EPMA   and    X-ray   diffraction   on   $T_c=75$~K   crystals,
(Tl$_{1.920(2)}$Cu$_{0.080(2)}$)Ba$_{1.96(2)}$CuO$_{6+\delta}$      and
(Tl$_{1.914(14)}$Cu$_{0.086(14)}$)Ba$_2$CuO$_{6.29(18)}$ respectively,
are closer to stoichiometric  than are typically reported on crystals,
possibly due to more effective retention of thallium.

\fussy
The trends in  the lattice parameters with doping  are consistent with
those   reported  previously   \cite{Wagner1997,Peets2007},   and  the
presence  of an orthorhombic  distortion in  overdoped superconducting
crystals is confirmed.  A  structure refinement indicated the presence
of a previously unobserved commensurate superlattice modulation in the
Ba--O and primarily Tl--O layers, doubling the unit cell along $a$ and
$b$, but not well correlated along $c$.  Further work will be required
to determine the exact nature of this modulation.

\section{Acknowledgements}

This  work  was supported  by  the  Natural  Sciences and  Engineering
Research Council of Canada (NSERC).  The authors are grateful to B. O.
Patrick  and  the  UBC  Vancouver  Chemistry  Department's  Structural
Chemistry Facility for assistance  with the collection and analysis of
X-ray diffraction data and to A. P. Mackenzie for helpful discussions.

\bibliographystyle{elsarticle-num}
\bibliography{dpeets-growth}
\end{document}